# Magneto-elastic coupling and unconventional magnetic ordering in multiferroic triangular lattice $AgCrS_2$

F. Damay[1], C. Martin[2], V. Hardy[2], G. André[1], S. Petit[1], and A. Maignan[2]

[1]Laboratoire Léon Brillouin, CEA-CNRS UMR 12, 91191 GIF-SUR-YVETTE CEDEX, France

[2]Laboratoire CRISMAT, CNRS UMR 6508, 6 bvd Maréchal Juin, 14050 CAEN CEDEX, France

AUTHOR EMAIL ADDRESS francoise.damay@cea.fr

ABSTRACT

The temperature evolution of the crystal and magnetic structures of ferroelectric sulfide $AgCrS_2$ have been investigated by means of neutron scattering. $AgCrS_2$ undergoes at $T_N$ = 41.6 K a first-order phase transition, from a paramagnetic rhombohedral *R*3*m* to an antiferromagnetic monoclinic structure with a polar *Cm* space group. In addition to being ferroelectric below $T_N$, the low temperature phase of $AgCrS_2$ exhibits an unconventional collinear magnetic structure that can be described as double ferromagnetic stripes coupled antiferromagnetically, with the magnetic moment of $Cr^{+3}$ oriented along *b* within the anisotropic triangular plane. The magnetic couplings stabilizing this structure are discussed using inelastic neutron scattering results. Ferroelectricity below $T_N$ in $AgCrS_2$ can possibly be explained in terms of atomic displacements at the magneto-elastic induced structural distortion. These results contrast with the behavior of the parent frustrated antiferromagnet and spin-driven ferroelectric $AgCrO_2$.

INTRODUCTION

Multiferroic materials, in which two or all three ferroic order parameters (ferroelectricity, (antiferro)magnetism, and ferroelasticity) are observed, have been the subject of intensive research in recent years. Such systems are rather rare in nature but are potentially interesting for a wide array of technological applications [1], [2]. Magnetic metal transition oxides with broken space-inversion symmetry are one of the new classes of magnetoelectrics being the most widely studied nowadays : in these so-called "spin driven ferroelectrics", it is the non-collinear spin spiral structure that is responsible for the inversion symmetry breaking [3], [4]. Examples of such oxides can be found amongst different structural families : perovskites ($TbMnO_3$ [3]), spinels ($CoCr_2O_4$ [5]) or delafossites (substituted $CuFeO_2$ [6], [7], $ACrO_2$ (A = Cu [8], [9], Ag [10]) to cite only a few. For instance, in $CuCrO_2$, a modulated helicoidal magnetic structure derived from the 120° arrangement expected on a 2D perfect triangular lattice is observed below $T_N$ = 24K [8], [11], and polarization in this compound probably arises [12] from a spin-orbit coupling induced modulation of the hybridization between the *3d* cations carrying the spin ($Cr^{+3}$) and the ligand oxygen ions. Shedding light on the specific role of the anions in these compounds is therefore a key issue. Structurally closely related to the $ACrO_2$ delafossites with its stacking of regular triangular layers, the $ACrS_2$ series is seemingly an ideal system to investigate this matter. In addition, a multiferroic ground state has been recently reported in $AgCrS_2$ [13]. In this context, the knowledge of the magnetic structure is of crucial importance, so as to validate the right physical model behind the multiferroic behavior. Surprisingly enough, though the magnetic structures of several $ACrS_2$ (A = $Cu^+$ [14], [15], $Na^+$ [16], $Tl^+$ [17]) compounds with $R3m$ symmetry have been investigated in the past, $AgCrS_2$ still remains to be studied. This motivated our neutron powder diffraction study of $AgCrS_2$ in the temperature range 1.5K-300K, which is reported hereafter. We show that, at the antiferromagnetic ordering transition, this compound undergoes a first-order phase transition to a monoclinic ferroelectric phase ; despite the stacked triangular layers topology of the compound, spins order collinearly within the layer planes, to form a stacking of double ferromagnetic stripes arranged antiferromagnetically (the so called 4-sublattice (4SL), or ↑↑↓↓ structure). Preliminary inelastic neutron scattering results are discussed in the light of the possible magnetic exchange paths in the distorted structure, to understand the magnetic couplings involved in the stabilization of this unconventional magnetic ordering.

## EXPERIMENTAL

5g of polycrystalline $AgCrS_2$ were prepared by high temperature solid state reaction. Powders of Ag, Cr and S precursors were weighted according to the stoichiometric ratio. The resulting powder was carefully ground and pressed in the shape of bars, and heated in an evacuated silica tube at 900°C for 12 hrs. The obtained sample was then checked by room temperature X-ray diffraction and found to be single phase.

Magnetic susceptibility defined as $\chi = M/H$ was calculated from magnetization data measured in a magnetic field of 0.1 Tesla, on warming from 1.5 to 300K, after a zero-field cooling, using a Quantum Design SQUID magnetometer. Heat capacity measurements were carried out in a Physical Property Measurement System (PPMS) using a semi-adiabatic relaxation method [18]. Outside the transition region, we used the standard $2\tau$ model [19] to fit at once the heating and cooling branches at each point. Around the transition, however, a single-pulse method (SPM) suited to first-order transitions has been used [20]. In the SPM, the temperature is swept across the complete width of the transition, and the heat capacity is derived from a point-by-point analysis of the time response along each of the two branches.

Neutron powder diffraction (NPD) versus temperature was performed on the G4.1 diffractometer ($\lambda$ = 2.425 Å) from 1.5 to 300K, and high resolution neutron diffractograms were recorded on the diffractometer 3T2 ($\lambda$ = 1.225Å) at 10K and 300K. Both diffractometers are located at LLB-Orphée (CEA-Saclay, France). Rietveld refinements and determination of the magnetic symmetry with representation analysis were performed with programs of the FullProf suite [21]. Inelastic neutron scattering experiments were performed on the thermal (2T, $k_f = 2.662$ Å$^{-1}$) and cold (4F2, $k_f = 2.662$ Å$^{-1}$ or 1.550 Å$^{-1}$) neutron triple-axis spectrometers at LLB-Orphée (Saclay, France). Higher order contaminations were removed with pyrolytic graphite or nitrogen-cooled Be filter placed in the scattered beam. Synchrotron diffractograms were recorded on the beam line I11 at the Diamond Light Source.

## RESULTS

### *Room temperature crystal structure*

In agreement with previous X-ray studies [16], [22], [23], the refinement of the room temperature neutron data (3T2) confirms that $AgCrS_2$ has a rhombohedral non-centrosymmetric $R3m$ crystal structure (Figure 1), with $a = b = 3.4979(1)$Å, and $c = 20.5369(9)$Å. This structure can be described as a stacking of edge-sharing octahedra layers $[CrS_2]_\infty$, connected through $AgS_4$ tetrahedra. Silver ions are ordered on half the tetrahedral sites, which form a pseudo two-dimensional puckered honeycomb lattice. Unlike in the delafossite $R\bar{3}m$ structure of $AgCrO_2$ [24], the symmetry of the $Cr^{+3}$ environment is not $D_{3d}$ (which corresponds to a distorted octahedron where all Cr-O distances are equal) but $C_{3v}$ : that is, a trigonal prism, with two sets of three equal Cr-S distances, 2.389(3)Å and 2.428(2)Å (inset of Figure 1b). The $AgS_4$ tetrahedron is also irregular, with one short (2.395(7)Å) and three long (2.723(4)Å) Ag-S distances (inset of Figure 1b). Within the regular triangular plane formed by the $Cr^{3+}$ lattice, the Cr-Cr distance is 3.4979(1)Å, which is much larger than in chromium oxides with delafossite structure (2.9843(4)Å in $AgCrO_2$ [25] for example) but similar to $CuCrS_2$ [23]. Results of the refinement, along with selected distances and angles, are summarized in Table I and III. The anisotropic displacement factors $U_{11}$ (displacement ellipsoid flattened along $c$) and $U_{33}$ (displacement ellipsoid elongated along $c$) reported in Table I indicate that the atomic motion of $Ag^+$ is strongly anisotropic and confined within the ($a$, $b$) plane, a frequent feature amongst layered compounds [26]. The $U_{11}$ displacement parameter of $Ag^+$ is actually exceptionally large ; this has been reported previously for $AgCrS_2$, in studies related to its superionic conductivity at high temperature [27], [28], [29], in particular, or $AgCrSe_2$ [16], [29], and to a lesser extend for $CuCrS_2$ [30]. In [29], silver ions were shown to be strongly involved in low frequency phonon modes corresponding to a vibration parallel to the $[CrS_2]_\infty$ layers. $AgCrS_2$ actually undergoes a reversible order-disorder ($R3m$ to $R\bar{3}m$) phase transition around $T_C$ = 670K [16], which corresponds to a statistical distribution of the Ag ions over all the tetrahedral positions between the $[CrS_2]_\infty$ layers in the high temperature phase. Interestingly, $R3m$ is a polar space group (along $[111]_R$, that is, along $c$ in the hexagonal cell), and ferroelectricity can be expected in $AgCrS_2$ below $T_C$. At room temperature, ferroelectricity is the result of the Cr plane being only slightly off-centered between the two sulfur planes, and should actually be very sensitive to the position of Ag ions along $c$. The atomic disordering of the $Ag^+$ cations is likely therefore, as mentioned in [29], to preclude the observation of an electric polarization at room temperature.

*Low temperature crystal structure*

The temperature dependence of the neutron diffractograms and of the susceptibility (Figure 2a and b) in the 300K-1.5K range shows the appearance of antiferromagnetic Bragg peaks, concomitantly with a sharp decrease of $\chi$, at $T_N$ = 41±1K, confirming the antiferromagnetic transition reported earlier [13], [31]. The Curie-Weiss temperature extracted from a Curie-Weiss law fit between 150K and 400K is in good agreement with previous results [16], [23], which report $\theta_{CW}$ = -55K, thus indicating predominantly antiferromagnetic interactions. The frustration parameter value $f = |\theta_{CW}|/T_N$ is about 1.3, which, like in $CuCrS_2$, is rather low for a triangular magnetic lattice [31]. The temperature evolution of the specific heat (Figure 2c) exhibits a very sharp peak (FWHM ~ 0.25K) at $T_N$, indicative of the first-order transition mentioned previously in [31]. This is further supported by the observation of a temperature shift of about 0.2K between the $T_N$ recorded upon cooling ($T_N$ = 41.5K) and that recorded upon warming ($T_N$ = 41.7K, not shown). The $Cr^{3+}$ ordered magnetic moment refined from the neutron data reaches ~2.7$\mu_B$, that is, 90% of its maximum value of 3$\mu_B$, within 4K (Figure 2d).

The transition to the antiferromagnetic state occurs simultaneously with a structural transition involving a discontinuous contraction of the distance between triangular planes (Figure 2e) and an anisotropic change of the Cr-Cr distances within the triangular plane (Figure 2f), resulting in a lowering of the symmetry to monoclinic *Cm* (Figure 3a and inset). *Cm* is a subgroup of *R*3*m*, but, in agreement with the specific heat measurements, the order parameter expansion [32] contains a third-degree invariant so that, under Landau condition [33], this phase transition should indeed be first order. The monoclinic cell parameters at 10K are $a$ = 13.7861(2) Å, $b$ = 3.5042(1) Å, $c$ = 7.1132(1) Å and $\beta$ = 155.276(5)°, the relationship between the rhombohedral and the monoclinic cells being illustrated on Figure 3b. The result of the refinement of the high-resolution neutron powder diffractogram of the *Cm* phase at 10K is illustrated on Figure 3a and summarized in Table II and III. Within the triangular plane, the distortion of the triangular lattice below $T_N$ (Figure 2f) leads to two long Cr-Cr distances along $[010]_m$ (3.5042(1) Å), and four short Cr-Cr distances of 3.4796(1) Å along $[½\ ½\ 1]_m$ and $[½\ \text{-}½\ 1]_m$ (Figure 3c). If we now consider second-neighbor distances within the triangular plane, the monoclinic distortion leads to a shorter Cr-Cr distance along the $[102]_m$ direction (6.0125(2) Å), compared to 6.0553(1) Å along $[1/2\ 3/2\ 1]_m$ and $[½\ \text{-}3/2\ 1]_m$. The Cr-Cr distances between triangular planes ($Cr\text{-}Cr_{out}$ in Table III) decrease slightly in the monoclinic phase compared to the rhombohedral one ; however the difference between the two sets of $Cr\text{-}Cr_{out}$ distances in the monoclinic phase lies within the experimental resolution, and their anisotropy as a result is far less pronounced than for the $Cr\text{-}Cr_{in}$ distances, whether first or second neighbor. It should also be pointed out here that the Cr-S distances are kept almost unchanged through the transition, the variations observed lying within the experimental error.

The $R3m \rightarrow Cm$ phase transition has been further studied in terms of atomic displacements or modes [34], to establish the instabilities at the origin of the distorted phase. Two macroscopic deformations are associated with the two irreducible representations of the space group $R3m$ for **k'** = (0 0 0) : a *c*-axis dilatation (or contraction), preserving the $R3m$ symmetry ($\Gamma_1$) and a shear deformation in the *ac* plane (the polar plane of the $Cm$ crystal class), which leads to the monoclinic $Cm$ symmetry (primary mode $\Gamma_3$). $\Gamma_1$ leads to a compression of the $AgS_4$ tetrahedron and to a slight expansion of the $[CrS_2]_\infty$ layer thickness. The effect of $\Gamma_3$ is a shearing of the $S_1$ and $S_2$ planes, which breaks the $C_{3v}$ symmetry of the $CrS_6$ octahedron. As a result of $\Gamma_3$, the polarization vector lies in the *ac* plane of the monoclinic cell, as there is an additional component of $\mathbf{P_S}$ parallel to the triangular plane. The amplitudes of these two modes are equivalent (0.0185Å for the primary mode and 0.0252Å for the secondary one) and rather large, almost comparable to those reported for archetypal ferroelectric materials [35] and could explain the appearance of a spontaneous polarization in $AgCrS_2$ in its monoclinic phase.

*Magnetic structure*

The magnetic Bragg peaks appearing below $T_N$ can be indexed with a commensurate propagation vector **k** = $(0\ 0\ ¼)_m$. Symmetry analysis [21] shows that the magnetic representation is decomposed into two irreducible representations, $\Gamma_{mag} = \Gamma_1 \oplus 2\Gamma_2$. The corresponding basis vectors are $\psi_1$ = (0 1 0) for $\Gamma_1$, and $\psi_{21}$ = (1 0 0) and $\psi_{22}$ = (0 0 1) for $\Gamma_2$. The only model compatible with the experimental data mixes two vectors transforming into $\Gamma_1$, $\psi_1 + i\psi_1$. The corresponding refinement (T = 1.5K), which leads to a satisfying magnetic Bragg agreement factor of 6.5%, is illustrated on Figure 4a. The magnetic structure is collinear, and can be described within the triangular plane as ferromagnetically aligned double spin stripes running along $b_m$, that are arranged antiferromagnetically (4SL structure) (Figure 4b). The $Cr^{+3}$ moment is oriented along $b_m$, and reaches 2.66(2)$\mu_B$ at 1.5K. The stacking of the magnetic planes is shifted so that $Cr^{+3}$ moments belonging to the $(001)_m$ plane are antiparallel along $a_m$ (Figure 4c). Diffraction data measured above $T_N$ also show a broad diffuse scattering signal around $Q_o$ = 0.6 $A^{-1}$ (inset of Figure 5), whose intensity gradually increases as the temperature nears $T_N$. As revealed by preliminary inelastic neutron scattering measurements, this diffuse scattering signal partly results from quasi-static fluctuations extending up to a few meV (Figure 5), as at energy transfers E = 3 meV and E = 0.4 meV, the inelastic spectra also exhibit a broad maximum around $Q_o$. Interestingly, this $Q_o$ value does not correspond to any of the magnetic Bragg peaks appearing below $T_N$ (Figure 5). To shed light on this issue, and assuming that the diffuse scattering I(Q) arises from correlations between nearest-neighbors spins following [36], we can write I(Q) as :

$$I(Q) = F(Q)^2 \sum_{i,j} \langle S_i S_j \rangle \frac{\sin Q\, r_{ij}}{Q r_{ij}}$$

where F(Q) is the Cr form factor, $S_i$ is the spin at site $i$ and $r_{ij}$ the distance between spins located on sites $i$ and $j$. The maximum at $Q_o$ yields a corresponding $r_{ij}$ = 6-7 Å, clearly pointing out that the magnetic correlations do not arise from nearest-neighbors interactions on the triangular lattice : in such a case, with $r_{ij}$ = 3.5 Å, we would expect I(Q) to be maximum around 1.25 $Å^{-1}$. Following this simple model, spin correlations seem rather unexpectedly to arise from second neighbors (~6.1Å, table III) and/or third neighbors (~6.9Å) interactions.

On cooling below $T_N$, the diffuse scattering signal vanishes as the magnetic Bragg peaks appear (inset of Figure 5). In parallel, the dynamic correlations at $Q_o$ abruptly disappear and a new inelastic response, probably spin-waves, emerges from the magnetic Bragg peaks (Figure 5). The existence of a spin gap cannot be inferred from the inelastic data : a weak signal is still detected at very low energies, showing that if there is a gap, its value is smaller than 0.4 meV. It seems therefore that $Cr^{3+}$ retains its Heisenberg character, like in $CuCrO_2$ [9]. The dispersion (not shown) is very steep, but further measurements, currently under progress, are required to obtain a more accurate picture of the spin dynamics.

DISCUSSION

The first order structural transition that occurs simultaneously with collinear antiferromagnetic ordering in $AgCrS_2$ is a strong indication of spin-lattice coupling, a behavior that was reported recently in $CuCrS_2$ as well [15]. The observation of a macroscopic polarization value at $T < T_N$ in this material stresses in addition that the electric dipole is coupled to the magnetic moment through the lattice. In this framework, the polarization measured by Singh et al. in $AgCrS_2$ seems to be reasonably interpreted on the basis of charges displacement at $T_N$, and more precisely to the shearing of the sulfur planes. To use the classification of multiferroics proposed by Cheong et al. [37], and in contrast to closely structurally related compounds like $AgCrO_2$ or $CuCrO_2$ [10], $AgCrS_2$ does not appear to belong to the class of spin-driven ferroelectrics, in which ferroelectricity results from non-collinear magnetic ordering. $AgCrS_2$ belongs rather to the geometric ferroelectric class [37], in which polarization becomes measurable following a lattice distortion induced by the magnetic ordering.

The magnetic structures of several $ACrS_2$ ($A = Li^+$, $Cu^+$, $Na^+$, $K^+$, $Tl^+$) compounds have been investigated in the past, and show a wide variety of magnetic arrangements : for the largest A radius, ferromagnetism has been reported ($A = Tl^+$ [17]), as well as ferromagnetic layers coupled antiferromagnetically ($A = K^+$ [38]). A 120° helicoidal magnetic structure is known for the smallest A ($A = Li^+$ [39]). Incommensurate helicoidal magnetic structures have also been reported for $A = Na^+$ [16] and $Cu^+$ [14], [15] ; $AgCrS_2$ seems therefore to be the first example in this family of the 4-sublattice antiferromagnetic structure.

The occurrence of an up-up-down-down structure on a triangular plane still remains puzzling. In a topologically rather similar compound, $CuFeO_2$, a $\uparrow\uparrow\downarrow\downarrow$ configuration has been reported below $T_N = 11K$, simultaneously with a first-order lattice distortion from $R\bar{3}m$ to $C2/m$ [40]. The magnetic exchanges that could stabilize such a structure are still hypothetical, and authors invoke a third neighbor interaction inside the triangular plane, in reference to the phase diagram of the 2D Ising spin on a triangular lattice [41]. The magnetic arrangement in the triangular plane is different (**k** = (0 ½ ½)), however, from the one observed in $AgCrS_2$, as the magnetic coupling along $b_m$ is antiferromagnetic.

The strong magneto-elastic coupling that is observed in $AgCrS_2$ could provide a way to understand the dominant magnetic exchanges stabilizing this complex magnetic structure: indeed, we observe that the monoclinic distortion induces two non-equivalent nearest neighbors couplings $J_1$ and $J'_1$, as well as two non-equivalent second neighbor couplings ($J_2$ and $J'_2$) (see the full and dotted lines on Figures 3, 4b and 4c). The observed 4-sublattice spin arrangement cancels the effect of $J'_1$, and $J'_2$, withdrawing the molecular field due to both couplings. We can therefore speculate that the relevant magnetic interactions are those that correspond to the remaining un-frustrated pathways, the ferromagnetic first neighbor coupling $J_1$ (along $b_m$) , the antiferromagnetic second neighbor super-super exchange $J_2$ (along Cr-$S_1$-$S_2$-Cr, perpendicularly to $b_m$), and the inter-plane antiferromagnetic super-super exchange $J_3$ (along $a_m$). Though the impact of the monoclinic distortion actually only subtly affects distances and angles, it is enough to lift the degeneracy between magnetic exchange paths and thus favors one magnetic configuration. Incidentally, a similar analysis holds in the case of the ↑↑↓↓ arrangement observed in $CuFeO_2$. Our inelastic scattering results also emphasize in addition the relevance of $J_2$ and/or $J_3$ couplings *above* $T_N$, as they correspond to interactions between Cr spins between 6 and 7 Å apart, that will give scattering around Q = 0.6$Å^{-1}$.

The fact that for large Cr-Cr distances ferromagnetic coupling is observed can be understood within the framework of Goodenough's model, which is based on a competition between direct cation-cation exchange and super cation-anion-cation exchange [42]. The direct Cr-Cr exchange across the common edges of adjacent $CrS_6$ octahedra involves half filled $t_{2g}$ orbitals of the $Cr^{3+}$ ions, and thus favors antiferromagnetic coupling. Super exchange through Cr-S-Cr with a close to 90° angle involves a half-filled $t_{2g}$ of a cation, an empty $e_g$ of the second cation, and an anionic *p* orbital, thus favoring ferromagnetic coupling. Direct-exchange is more sensitive to the distance between cations than super-exchange, and for a threshold value of the Cr-Cr distance, indirect exchange will start to predominate, thus leading to a ferromagnetic arrangement of the Cr moments. Rosenberg *et al.* [17] evaluated this threshold value to be ~ 3.6Å : in $AgCrS_2$, ferromagnetic spin arrangement is observed for Cr-Cr = 3.5042(10)Å. The role of the ligand anion on the magnetic couplings in this material is still a matter of speculations, but it should provide a most interesting field of future research, both on experimental and theoretical grounds.

## CONCLUSION

The investigation of $AgCrS_2$ by neutron scattering techniques has evidenced a magnetoelastic coupling induced structural transition at $T_N$, towards a polar monoclinic phase. The antiferromagnetic arrangement is rather unusual compared to $CuCrS_2$ or to frustrated parent delafossite compounds like $AgCrO_2$, and can be described as double ferromagnetic stripes coupled antiferromagnetically, running along $[010]_m$. How this magnetic structure is stabilized can be understood using a ferromagnetic coupling along $b_m$ - the direction which corresponds to the largest Cr-Cr distance in the anisotropic triangular lattice -, in addition to an antiferromagnetic second neighbor coupling perpendicular to $b_m$, that could be the driving force behind the structural distortion.

## ACKNOWLEDGMENTS

Authors acknowledge F. Porcher for neutron diffraction experiments on 3T2 (LLB-Orphée). Part of this work was also carried out with the support of the Diamond Light Source and we thank Dr. C. Tang and S. Thompson for their help during the experiment. Financial support for this work was partially provided by the French Agence Nationale de la Recherche, Grant No ANR-08-BLAN-0005-01.

TABLE CAPTIONS

Table I. Rietveld refinement results of high resolution neutron powder diffractogram of $AgCrS_2$ at 300K (space group *R3m* (n°160, H setting) with all atoms on Wyckoff position 3*a* (0, 0, *z*)).

| | |
|---|---|
| Temperature | 300K |
| Space group | *R3m* (n° 160) |
| Cell parameters (Å) | |
| *a* | 3.4979(1) |
| *b* | 3.4979(1) |
| *c* | 20.5369(9) |
| Cell volume *V* (Å$^3$) | 217.6(4) |
| Ag (0, 0, *z*) | 0.1545(2) |
| Cr (0, 0, *z*) | 0 |
| $S_1$ (0, 0, *z*) | 0.2712(3) |
| $S_2$ (0, 0, *z*) | 0.7323(2) |
| $U_{anisotropic}$ ($10^{-4}$ Å$^2$) | |
| Ag $U_{11}$ | 534(9) |
| Ag $U_{33}$ | 133(11) |
| Cr $U_{11}$ | 51(4) |
| Cr $U_{33}$ | 107(12) |
| $S_{1,2}$ $U_{11}$ | 64(3) |
| $S_{1,2}$ $U_{33}$ | 76(10) |
| Number of reflections | 107 |
| Number of parameters | 16 |
| Bragg R factor | 3.21 |
| $\chi^2$ | 1.26 |

Table II. Rietveld refinement results of high resolution neutron powder diffractogram of $AgCrS_2$ at 10K (space group *Cm* (n°8) with all atoms on Wyckoff position 2a ($x$, 0, $z$)).

| | |
|---|---|
| Temperature | 10K |
| Space group | *Cm* (n°8) |
| Cell parameters (Å) | |
| $a$ | 13.7861(2) |
| $b$ | 3.5042(1) |
| $c$ | 7.1132(1) |
| $\beta$ (°) | 155.276(5) |
| Cell volume $V$ (Å$^3$) | 143.7(4) |
| Ag $x$ | 0.8459(8) |
| $z$ | 0.1524(16) |
| Cr $x$ | 0 |
| $z$ | 0 |
| $S_1$ $x$ | 0.7263(10) |
| $z$ | 0.2660(20) |
| $S_2$ $x$ | 0.2688(12) |
| $z$ | 0.7364(24) |
| $U_{isotropic}$ ($10^{-4}$ Å$^2$) | |
| Ag | 101(5) |
| Cr | 75(6) |
| $S_{1,2}$ | 73(7) |
| Number of reflections | 2311 |
| Number of parameters | 20 |
| Bragg R factor | 3.66 |
| $\chi^2$ | 2.27 |

Table III. Selected inter-atomic distances and angles in $AgCrS_2$ at 300K and 10K (from high resolution neutron powder diffractograms).

| Temperature | 300K | 10K |
|---|---|---|
| Distances (Å) | | |
| Ag-$S_1$ | 2.395(7) | 2.41(7) |
| Ag-$S_2$ | 2.723(4) x3 | 2.70(6) |
| | | 2.70(5) x2 |
| Cr-$S_1$ | 2.389(3) x3 | 2.39(3) |
| | | 2.38(3) x2 |
| Cr-$S_2$ | 2.428(2) x3 | 2.43(3) |
| | | 2.42(4) x2 |
| Cr-$Cr_{in}$ | 3.4979(4) x6 | 3.5042(10) x2 |
| | | 3.4796(10) x4 |
| Cr-$Cr_{out}$ | 7.1373(1) x6 | 7.1131(5) x2 |
| | | 7.1122(2) x4 |
| Cr-$Cr_{in}$ ($2^{nd}$ neighbour) | 6.0585(7) x6 | 6.0125(2) x2 |
| | | 6.0553(1) x4 |
| $CrS_2$ layer thickness (Å) | 2.624(6) | 2.63(2) |
| $AgS_4$ height (Å) | 4.222(6) | 4.19(2) |
| Angles (°) | | |
| S-Cr-$S_{in}$ | 92.16(8) | 92.51(4) |
| S-Cr-$S_{out}$ | 86.84(1) | 87.24(4) |
| S-Ag-S | 132.11(1) | 131.8(5) |

FIGURE CAPTIONS

Figure 1 (color online) : (a) *R*3*m* crystal structure of $AgCrS_2$. (b) Refinement of the 3T2 neutron powder diffraction diffractogram of $AgCrS_2$ at 300K (experimental data : open circles, calculated profile : continuous line, allowed Bragg reflections : vertical marks. The difference between the experimental and calculated profiles is displayed at the bottom of the graph). Inset : environments of $Cr^{+3}$ and $Ag^{+}$.

Figure 2 (color online) : Temperature evolution of the G4.1 neutron diffractograms (a), of the zero-field cooling susceptibility in 0.1T (from reference [13]) (b), of the specific heat (c), of the refined magnetic moment (d), of the distance between triangular planes $d_{inter}$ (e) and of the Cr-Cr distances within the $[CrS_2]_\infty$ layer (f) of $AgCrS_2$. Lines are guide to the eye.

Figure 3 (color online) : (a) Refinement of the 3T2 neutron powder diffraction diffractogram of $AgCrS_2$ at 10K (experimental data : open circles, calculated profile : continuous line, allowed Bragg reflections : vertical marks. The difference between the experimental and calculated profiles is displayed at the bottom of the graph). Inset : synchrotron X-ray data at 300K (blue) and 15K (red) evidencing the monoclinic structural distortion. (b) Relationship between the rhombohedral (thin grey lines) and the monoclinic cells (thick red lines). (c) Distances (in Å) at 10K in the triangular plane. The cell distortion has been emphasized.

Figure 4 (color online) : (a) Refinement of the G4.1 neutron diffraction data at 1.5K (experimental data : open circles, calculated profile : continuous line, allowed Bragg reflections : vertical marks. The difference between the experimental and calculated profiles is displayed at the bottom of the graph). Magnetic peak indexation **H**+**k** is shown for the two main magnetic reflections. (b) Magnetic arrangement within the $Cr^{+3}$ triangular lattice. (c) Projection of the magnetic structure along $b_m$ (top) and along $[101]_m$ (bottom) illustrating the magnetic planes stacking (+ and – signs refer to spins belonging to the triangular plane directly above or underneath the reference one, respectively). The magnetic cell is shaded in grey, the crystal unit cell in red. In (b) and (c) dotted lines show degenerate directions along which the ↑↑↓↓ configuration is found. Full lines correspond to ferromagnetic (red lines) or in-plane (green) and inter-plane (blue) antiferromagnetic spin configuration. The color scheme is the same as in Figure 3c.

Figure 5 (color online) : Temperature dependence of the inelastic scattering profiles at E = 0.4 meV ($k_f$ = 1.55Å$^{-1}$) of $AgCrS_2$. Inset : Temperature evolution of the diffraction profiles (G4.1 data) emphasizing the diffuse magnetic scattering around 0.6Å$^{-1}$ just above $T_N$. On both graphs the dotted red lines indicate the positions of the two magnetic Bragg peaks observed in this Q range below $T_N$.

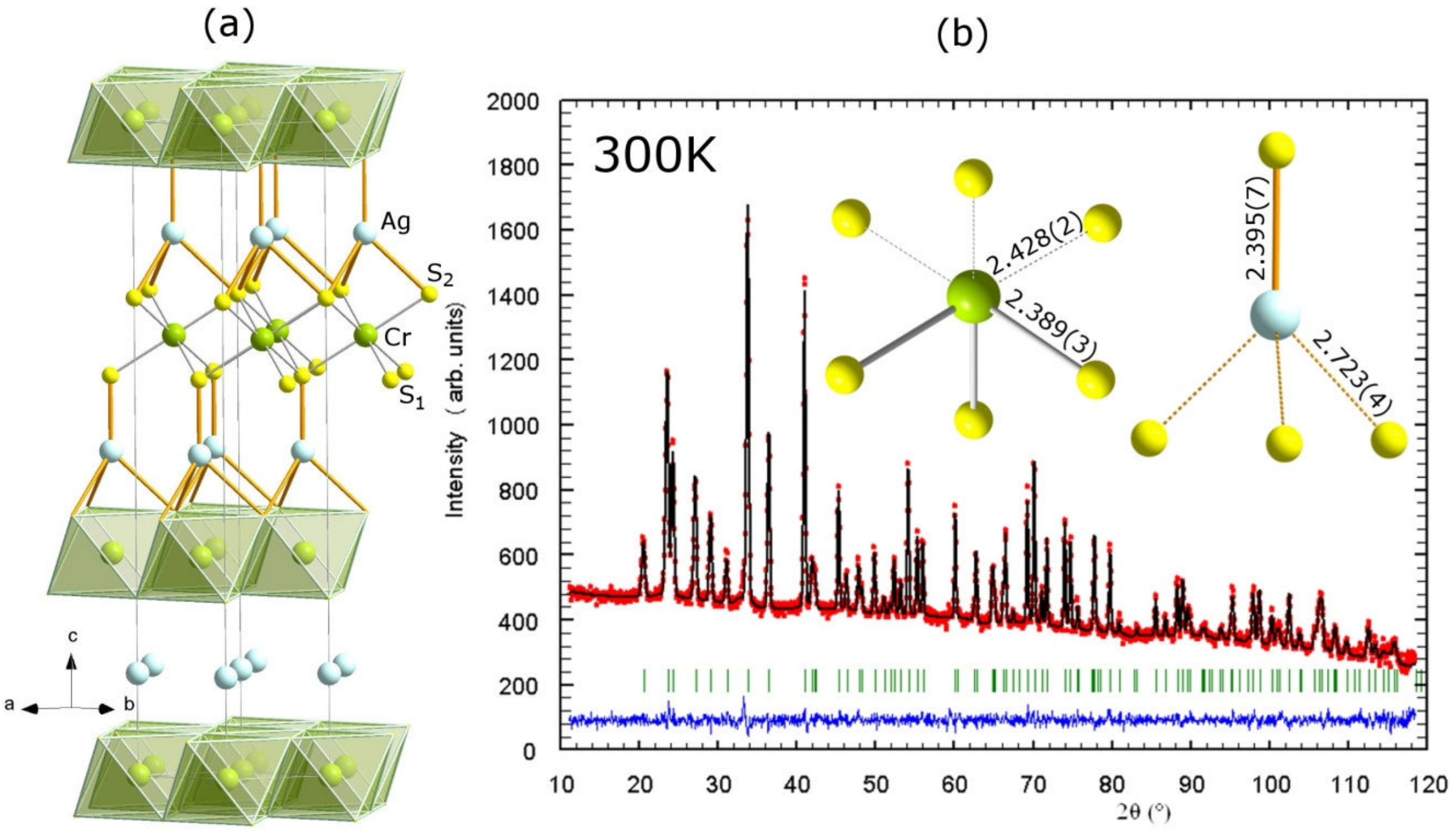

Figure 1

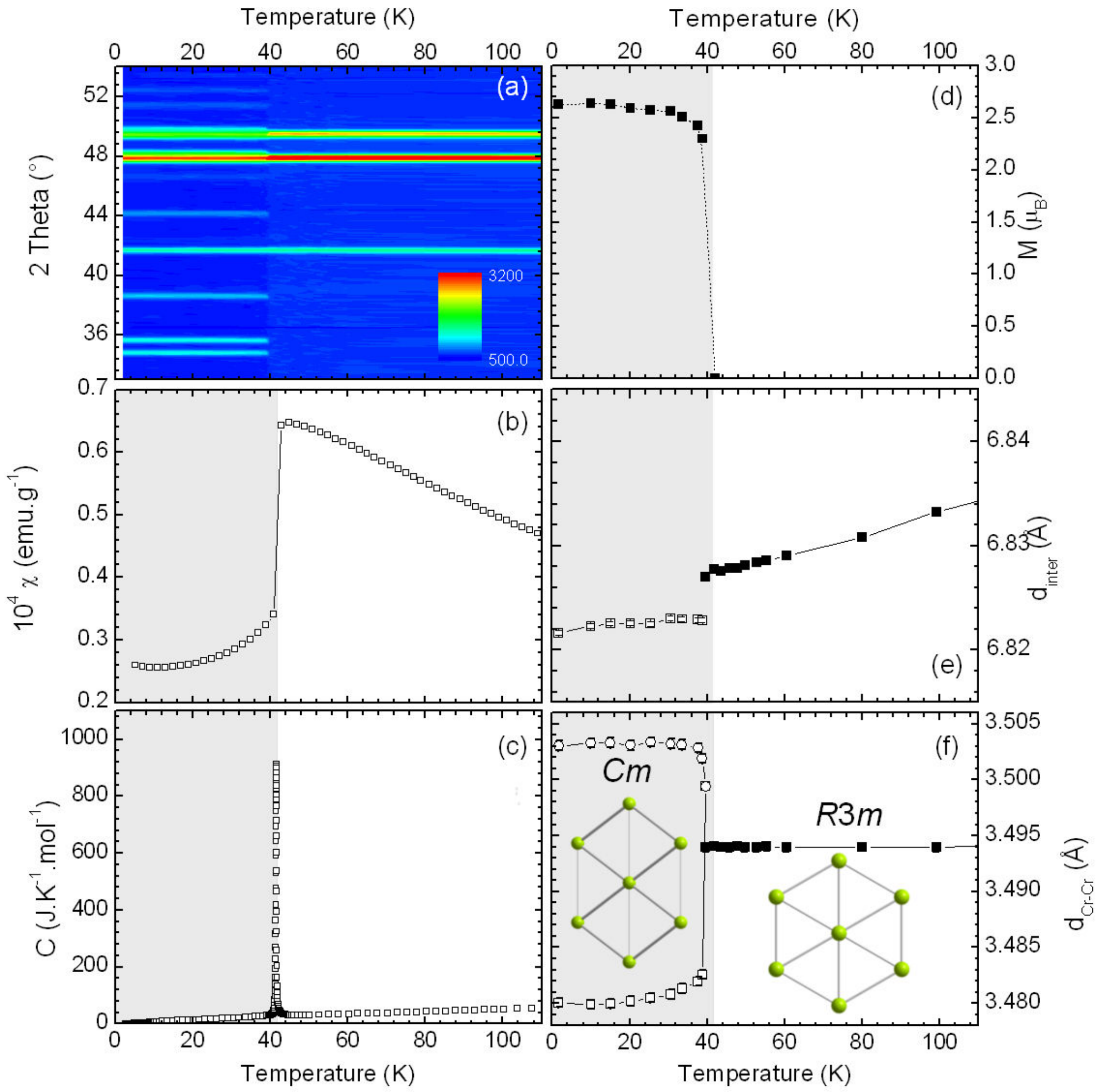

Figure 2

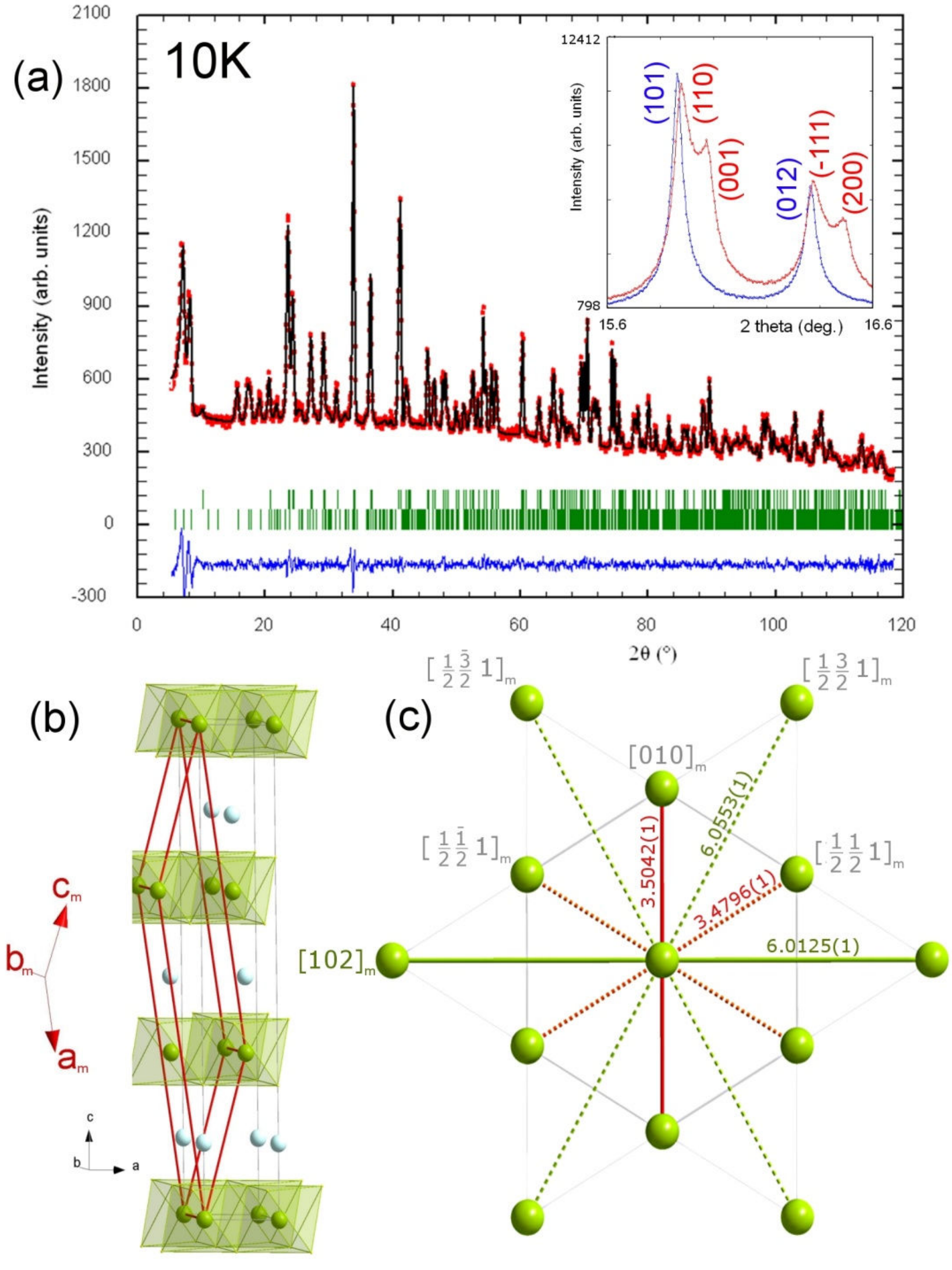

Figure 3

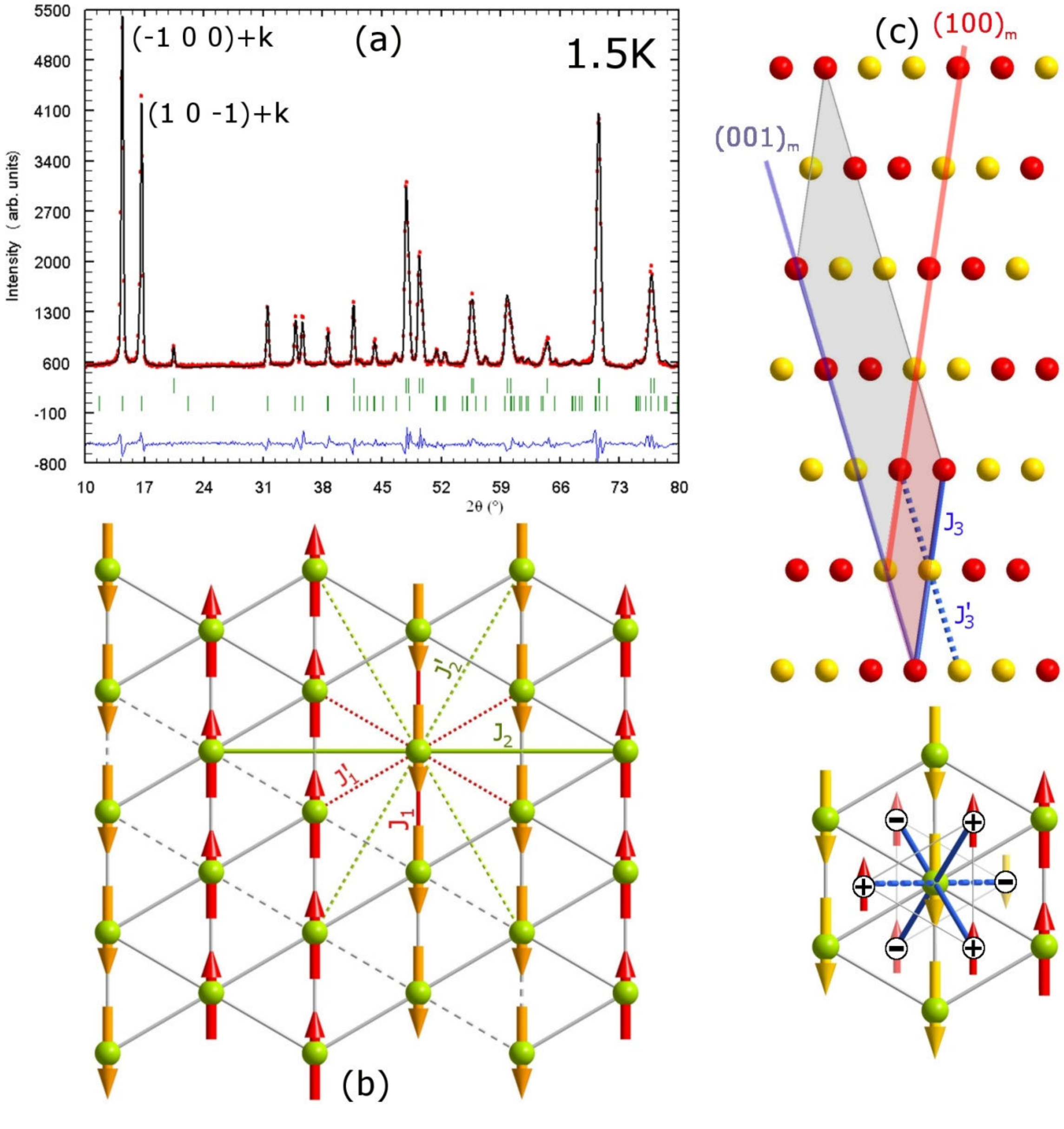

Figure 4

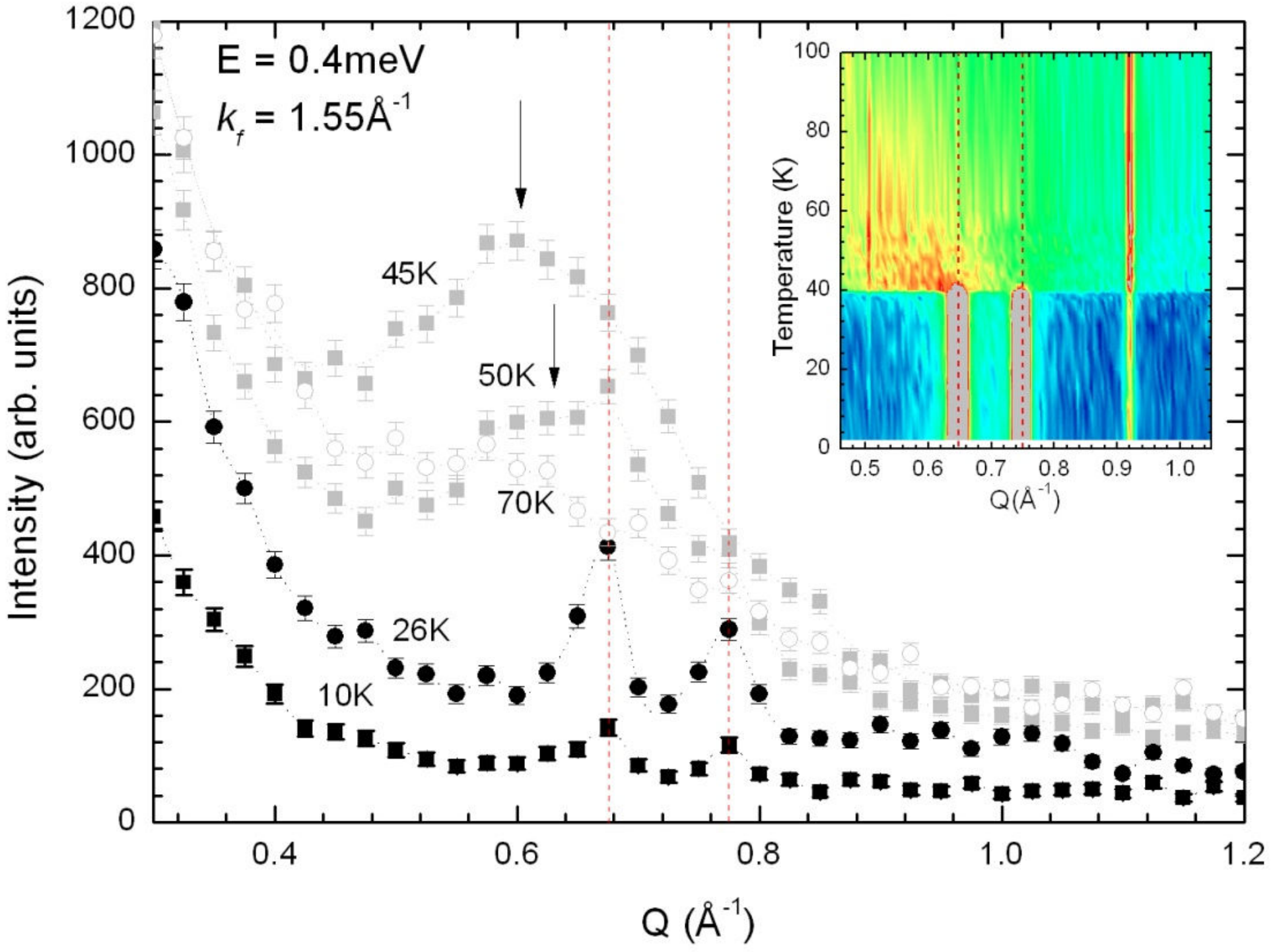

Figure 5